\documentclass[11pt]{article}
\usepackage{amsmath,amssymb,color,graphics,epsfig,cite}
\usepackage{hyperref}
\usepackage{ulem}
\usepackage{graphicx} 
\usepackage{tikz}
\usepackage{array}

\usepackage{amsfonts}

\newcommand{\be}{\begin{equation}}
\newcommand{\ee}{\end{equation}}
\newcommand{\bea}{\setlength\arraycolsep{2pt} \begin{eqnarray}}
\newcommand{\eea}{\end{eqnarray}}
\newcommand{\nn}{\nonumber}

\def\ft#1#2{{\textstyle{\frac{\scriptstyle #1}{\scriptstyle #2} } }}
\def\fft#1#2{{\frac{#1}{#2}}}

\def\0{{\sst{(0)}}}
\def\1{{\sst{(1)}}}
\def\2{{\sst{(2)}}}
\def\3{{\sst{(3)}}}
\def\4{{\sst{(4)}}}
\def\5{{\sst{(5)}}}
\def\6{{\sst{(6)}}}
\def\7{{\sst{(7)}}}
\def\8{{\sst{(8)}}}
\def\sst#1{{\scriptscriptstyle #1}}

\begin{document}

 
\begin{center}
		{\Large {\bf Regular black holes and their singular families}}

		\vspace{40pt}
  
{\large Hyat Huang\footnote{hyat@mail.bnu.edu.cn;} and Xiao-Pin Rao\footnote{xiaopingrao@jxnu.edu.cn;}}

\vspace{10pt}

{\it College of Physics and Communication Electronics, \\
	Jiangxi Normal University, Nanchang 330022, China}

\vspace{40pt}
		
		\underline{ABSTRACT}
	\end{center}

Regular black holes without curvature singularity can arise in Einstein gravity with appropriate matter energy-momentum tensor. We show that these regular solutions represent only a special case of a much broader family of black holes with a free mass parameter. The regularity is achieved only at a specific mass value, and any deviation from the fine-tuned parameter inevitably results in curvature singularity. As a concrete example, we consider nonlinear electrodynamics (NLED) as matter sources. A new NLED theory is proposed that is a generalization of the Bardeen class and the Hayward class. New regular black holes and their singular counterparts are obtained. Significant distinctions between regular black holes and their singular counterparts are analyzed. These findings provide new insights into regular black holes.

	\thispagestyle{empty}
	
\addtocontents{toc}{\protect\setcounter{tocdepth}{2}}
	

 
 \newpage

\section{Introduction}

Is there a singularity inside a black hole horizon? The answer would normally be “yes”. Black hole as a solution of the vacuum Einstein equations, such as the Schwarzschild black hole or the Kerr black hole, contains a singularity. R. Penrose suggested that these singularities are a consequence of gravitational collapse and cannot be avoided \cite{Penrose:1964wq}. However, there is another type of black holes that has a horizon without any curvature singularity. Such black holes go by the name of “regular black holes”. 

The first regular black hole model was proposed by Bardeen in 1968 \cite{bardeen}. After that, many regular black holes were found \cite{Hayward:2005gi,Ayon-Beato:1998hmi,Ayon-Beato:2004ywd,Rodrigues:2018bdc,Hu:2023iuw,Lan:2023cvz,Ying:2022xaj,Ying:2023gmd}. To obtain such solutions within General Relativity, one has to introduce exotic matter sources. For example, Ayón-Beato and García presented a non-linear electrodynamics (NLED) theory that admits the Bardeen black hole as a solution \cite{Ayon-Beato:2000mjt}. Afterwards, many works showed that regular black holes can be solutions of NLED theories \cite{Bronnikov:2000vy,Fan:2016hvf,Balart:2014cga,Nojiri:2017kex,Li:2023yyw,Li:2024rbw}. Although a recent study doubts the stability of regular black holes in NLED \cite{DeFelice:2024seu}. There are also other exotic matter sources, such as phantom scalars\cite{Bronnikov:2005gm,Bronnikov:2012ch} and Yang-Mills field\cite{Lavrelashvili:1992ia}, that can generate regular black hole solutions. Or within the framework of modified gravities, regular black hole solutions exist \cite{Berej:2006cc,Junior:2015fya,Rodrigues:2015ayd,Ghosh:2018bxg,Ashtekar:2023cod,deSousaSilva:2018kkt,Bueno:2024dgm}.
Thus, it is possible to remove singularities inside black hole horizons if we release the requirements on matter sources or the gravity theories themselves. The interior of a black hole horizon can be a (A)de Sitter/Minkowski core \cite{Lemos:2011dq,Simpson:2019mud}, a wormhole throat \cite{Nojiri:2024dde} or a black bounce \cite{Simpson:2018tsi,Huang:2019arj,Franzin:2021vnj}, etc. 

However, a question arises: do the theories admit only regular black holes? In this work, we aim to address this question in the static and spherically symmetric background within the framework of General Relativity.  Based on the Einstein equations, we show that if a theory admits a regular black hole solution, it will also have a singular one. For example, the Bardeen black hole has no free parameters in the NLED theory, namely, the black hole mass and magnetic charge are fixed by the theoretical parameters. The full solution of Bardeen black hole was found in Ref.\cite{Fan:2016hvf} and contains singularity.  We show that this phenomenon occurs not only in NLED theories but also with any matter sources. It proves that the regular black holes are merely fine-tuned objects in General Relativity. 

The rest of the paper is organized as follows. In Sec. \ref{2}, we demonstrate that regular black holes are fine-tuned solutions in General Relativity with arbitrary matter sources.  As an example, we consider an NLED theory and obtain a new black hole solution in Sec. \ref{3}. In Sec. \ref{4}, we investigate distinct features between regular black holes and their singular counterparts. Finally, we conclude our work in Sec. \ref{5}.

\section{Regular black holes as fined-tuned solutions}\label{2}

Historically, regular black holes are designed metrics without concrete theories. Later, one found most of them can arise in General Relativity with exotic matter sources. However, once fixing the theoretical parameters of theories, there is \textit{no} any free parameter in regular black hole solutions. It implies the most general solution to the theories is not a regular black hole. We may draw a conclusion: \textit{regular black holes are fine-tuned objects}.

To proof this statement, let's consider a general matter theory in the Einstein gravity. The Einstein equation is given by 
\bea\label{eom}
G_{\mu\nu} \equiv \mathcal{R}_{\mu\nu}-\fft{1}{2} \mathcal{R} g_{\mu\nu}=\kappa^2 T_{\mu\nu}.
\eea
Without loss the generality we set $\kappa=1$ in the rest of the paper. If one wants to read off the matter energy-momentum tensor $T_{\mu\nu}$ from the metric, one needs to be careful, since the components depend on the coordinate choice. In order to obtain a result that is independent of the local coordinate choice, one typically uses the vielbein basis
\be
T_{\nu}^{\mu}=\{-\rho, p_r, p_T,p_T\}.
\ee
For diagonal metrics, we can chose the vielbein base that corresponds to $\eta_{ab}$, then
\be
G^{a}_{b}=T^{a}_{b}=G^{\mu}_{\nu}=T^{\mu}_{\nu}.
\ee
The energy-momentum tensor is then diagonal in vielbein base with $\{-\rho,\Vec{p}\}$. Therefore we have
\be
\rho=-G^{t}_{t},\quad p_r=G^{r}_{r},\quad p_T=G^{\theta}_{\theta},
\ee
for a spherically symmetric and static metric
\be\label{ans}
ds^2=-h(r)dt^2+f(r)^{-1}dr^2+r^2d\Omega_2^2,
\ee
where $d\Omega^2_2$ is an unite $2$ sphere. We have
\bea\label{em}
\rho &=&-\fft{rf'+f-1}{r^2},\qquad p_r=\fft{f rh'+(f-1)h}{hr^2}\,,
\nn\\
p_T&=&\fft{2h^2f'+h(rf'h'+2f(rh''+h'))-fr(h')^2}{4h^2r}.
\eea
Now we assume there is a regular black hole solution $(h_0,f_0)$ satisfies the  energy-momentum tensor given in \eqref{em} and with $g_{tt}=-\ft{1}{g_{rr}}$. Thus we replace $(h,f)$ with $(f_0, f_0)$. We denote this energy-momentum tensor as $T^{(0)}_{\mu\nu}$. To obtain the most general solution of the energy-momentum tensor $T^{(0)}_{\mu\nu}$, we consider a deviation of the regular black hole
\be
g_{tt}=f=f_0+\tilde{f}, \qquad
g_{rr}^{-1}=f_0+\tilde{h}.
\ee
The Einstein equation now is 
\be
E^{\mu}_{\nu}=G^{\mu}_{\nu}-T^{(0)\mu}_{~~~\nu}=0.
\ee
The $E^t_t-E^r_r=0$ equation implies $\tilde{h}= \tilde{f}$. Then the remaining equation is 
\be
E^t_t=0,\quad \to \quad r\tilde{f}'+\tilde{f}=0.
\ee
Integrating this first-order equation gives rise to 
\be
\tilde{f}=-\ft{m}{r},
\ee
where $m$ is a free constant. 

Consequently, the full solution of the theory is given by 
\be
f(r)=f_0(r)-\ft{m}{r},
\ee
where $f_0$ depicts a regular black hole. The additional term $-\ft{m}{r}$ causes a curvature singularity at $r\to 0$.
Furthermore, this term may also contribute to the mass of the solution. We will show this fact in a concrete example in the next section.

The result can directly generalize to regular black holes with asymptotic (A)dS boundary.

\section{Examples: black holes in NLED}\label{3}

We consider a theory that has the Lagrangian
\be\label{lagBar}
{\cal L}=\sqrt{-g}(R-{{\cal F}}(F^2))\footnote{Where $F=dA$ and $F^2=F_{\mu\nu}F^{\mu\nu}$.},
\ee
with the NLED term 
\be\label{FNLED}
{\cal F}(F^2)=\ft{1 }{ \alpha}\ft{ (\alpha F^2)^\gamma}{(1+(\alpha F^2)^\beta)^\sigma}.
\ee 
There are four theoretical parameters $(\alpha, \beta, \gamma,\sigma)$ in the theory. The variations of metric $g_{\mu\nu}$ and gauge field $A_\mu$ give rise to the E.O.Ms
\bea
&&G_{\mu\nu}=R_{\mu\nu}-\ft{1}{2}g_{\mu\nu}R=2\ft{\partial {\cal F}}{\partial F^2} F_{\mu\nu}^2-\ft{1}{2}g_{\mu\nu}{\cal F}(F^2),\label{Geom}  ~\\ 
&&\nabla_\mu(\ft{\partial {\cal F}}{\partial F^2} F^{\mu\nu})=0 .\label{Meom}
\eea
We take the static and spherically symmetric ansatz
\be
ds^2=-f(r) dt^2+f(r)^{-1}dr^2+r^2d\Omega_2^2
\ee
for the metric, and a pure magnetic field ansatz
\be
F_{\theta \phi}=-F_{\phi\theta }=2p\sin \theta,
\ee
for the gauge field.
It can be verified that if $p$ is a constant then it is a solution of \eqref{Meom}. The physical meaning of $p$ is the magnetic charge.
Then the non-vanished Einstein equation \eqref{Geom} reduces to a first order differential equation
\be
-1+2^{3\gamma-1}p^{2 \gamma} r^{2-4\gamma}\alpha^{-1+\gamma}(1+8^{\beta}p^{2\beta}r^{-4\beta}\alpha^{\beta})^{-\sigma}+h(r)+r h'(r)=0.
\ee
By solving the above equation directly, we obtain the metric function
\be\label{solgeneral}
f(r)=1-\ft{m}{r}+\ft{1}{-3+4\gamma}2^{3\gamma-1}p^{2\gamma}r^{2-4\gamma}\alpha^{-1+\gamma} {_2F_1[\ft{-3+4\gamma}{4\beta},\sigma,\ft{-3+4\beta+4\gamma}{4\beta},-8^{\beta} p^{2\beta}r^{-4 \beta} \alpha^{\beta}]}.
\ee
There are two integration constants $(m, p)$ in the solution. The magnetic charge $p$ comes from the matter source while the parameter $m$ origins from gravity.

\subsection{Reduce to the Bardeen class}

In 1968, Bardeen presented a regular black hole metric and later Ayón-Beato and García found it is a solution to an NLED theory. The Lagrangian is given by 
\be\label{FB}
{\cal F}_B(F^2)=\ft{1}{\alpha}\ft{(\alpha F^2)^\fft{5}{4}}{(1+\sqrt{\alpha F^2})^\fft{5}{2}}.
\ee
This theory is a special case of our NLED theory, which relates to \eqref{FNLED} with $\gamma=\ft{5}{4}, \beta=\ft{1}{2}, \sigma=\ft{5}{2}$.
This theory admits regular black hole solution, namely the Bardeen solution, which is given by 
\be\label{barsol}
f_B(r)=1-\ft{r^2}{6\alpha (1+r^2)^{\fft{3}{2}}}, \qquad F_{\theta\phi}=\ft{1}{\sqrt{2\alpha}}\sin\theta.
\ee
Note that this solution is a fine-tuned case of $m=0$ and $p=\ft{1}{2\sqrt{2\alpha}}$ of \eqref{solgeneral} with the same setting of other parameters. Hence we know the full solution to this case is
\be \label{fullsol}
f(r)=1-\ft{2^{\fft{7}{4}}p^{\fft{3}{2}}r^2}{3\sqrt{2}(r^2+2\sqrt{2\alpha}p)^{\fft{3}{2}}\alpha^{\fft{1}{4}}}-\ft{m}{r}, \qquad F_{\theta\phi}=2p \sin\theta.
\ee
This two parameter solution \eqref{fullsol} was firstly found in Ref.\cite{Fan:2016hvf} and it contains a singularity at $r=0$ when $m\neq 0$.

\subsection{Reduce to the Hayward class}

The NLED theory \eqref{FNLED} and solution \eqref{solgeneral} can also reduce to another well-known regular black hole class, namely the Hayward class. The Lagrangian of Hayward class is given by
\be\label{FH}
{\cal F}_H(F^2)=\ft{1}{\alpha} \ft{(\alpha F^2)^\fft{3}{2}}{(1+(\alpha F^2)^\fft{3}{4})^2}.
\ee
With the setting $\gamma=\ft{3}{2}, \beta=\ft{3}{4}, \sigma=2$, our NLED theory \eqref{FNLED} reduces \eqref{FH}. It admits a regular solution, namely the Hayward black hole
\be
f_H(r)=1-\ft{r^2}{6\alpha(1+r^3)}, \qquad F_{\theta\phi}=\ft{1}{\sqrt{2\alpha}}\sin\theta.
\ee
Similarly to the Bardeen solution, the Hayward solution has no free parameters.
The corresponding full solution with two free parameters $(m,p)$ is given by 
\be
f(r)=1+\ft{8\sqrt{2\alpha }p^3}{3(r^4+2^{\fft{9}{4}}p^{\fft{3}{2}}\alpha^{\fft{3}{4}}r)}-\ft{m}{r}, \qquad F_{\theta\phi}=2p \sin\theta,
\ee
which reduces to the regular solution when $m=\ft{1}{6\alpha}$ and $p=\ft{1}{2\sqrt{2\alpha}}$. This full solution was found in Ref.\cite{Fan:2016hvf} firstly.

\section{New features in the singular families}\label{4}

Firstly, a regular black hole with a de Sitter core typically has two horizons. A singularity arises in full solution. From a mathematical perspective, there are two possible scenarios: if the singularity is time-like, the solution shows either a two-horizon black hole or a naked singularity, which would not differ significantly from the regular solution.

However, if the singularity is space-like, many interesting effects appear. Since the spacetime is asymptotically Minkowski, there is always at least one horizon. In the full solutions of the Bardeen and Hayward models, there can be up to three horizons. As the parameter $m$ changes, any two of these horizons can merge to form an extremal horizon.

We take the Bardeen class as an example to show these features. One should note that the Hayward class shares similar properties.

\subsection{Extra horizons form when regular solutions evolve to their singular counterparts}

One expands the metric function $f(r)$ in \eqref{fullsol} at $r\to 0$ gives rise to 
\be
f(r\to 0)=1-\ft{m}{r}-\ft{ r^2}{6\alpha}+\ft{r^4}{8\sqrt{2} p \alpha^\fft{3}{2}}\cdots.
\ee
It shows that there is a space-like and a time-like singularity arises when $m>0$ and $m<0$, respectively. When $m=0$, $r\to 0$ is finite and as a horizonless de Sitter core. 

The asymptotic behavior of $f(r)$ at large $r$ is given by
\be
f(r\to\infty)=1-\ft{1}{r}(m+\ft{2^{\fft{5}{4}} p^{\fft{3}{2}}}{3 \alpha^{\fft{1}{4}}})+\ft{2^{\fft{7}{4}}p^{\fft{5}{2}}\alpha^{\fft{1}{4}}}{r^3}+\cdots.
\ee
The mass of the solution can be read as $M=\ft{m}{2}+ \ft{2^{\fft{1}{4}}p^{\fft{3}{2}}}{3  \alpha^{\fft{1}{4}}}$. To make sure the mass is positive, one requres $m>-\ft{2^{\fft{5}{4}}p^{\fft{3}{2}}}{3  \alpha^{\fft{1}{4}}}$. 

\begin{figure}[t]
\centering
\includegraphics[width=0.45\textwidth]{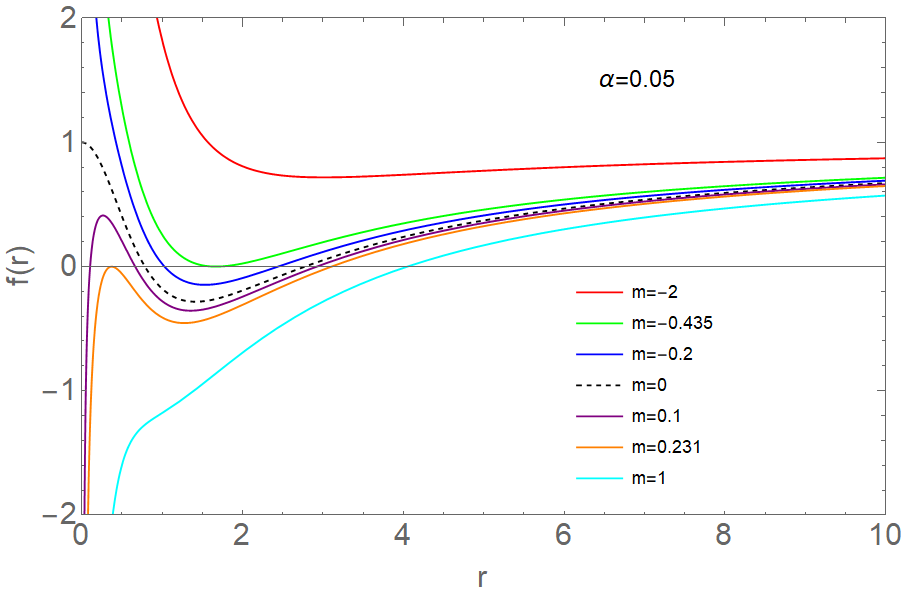} \qquad
\includegraphics[width=0.45\textwidth]{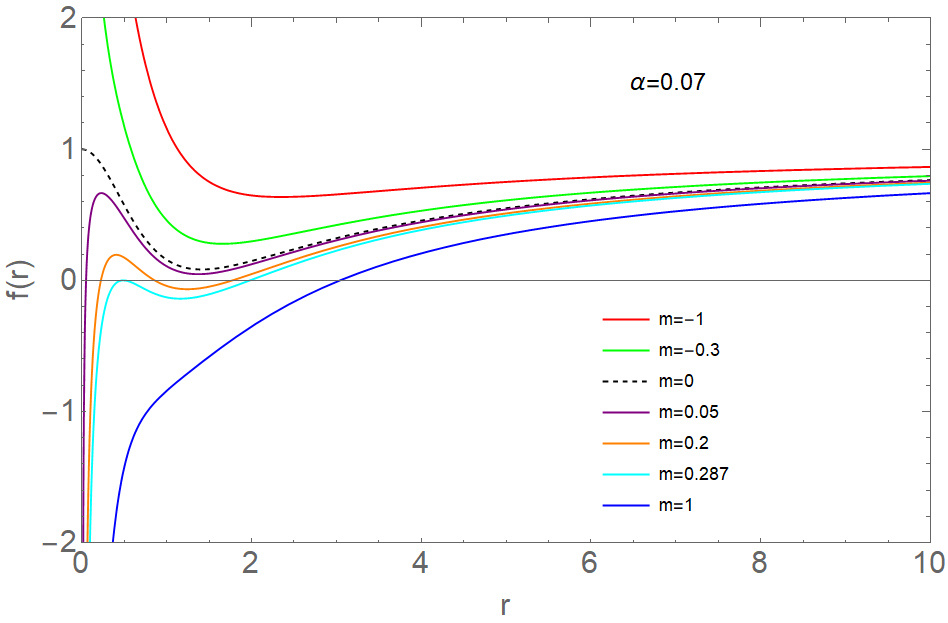}
\caption{ \it We show how the metric function change from a regular solution by varying its parameter $m$. We set $p=\ft{1}{2\sqrt{2\alpha}}$. The initial regular solution describes a two-horizon black hole without singularity (left plot) or a horizonless de Sitter core (right plot) in black dashed lines.
With the decreasing of $m$ from $0$, a time-like singularity arises at $r\to 0$. With the increasing of $m$ from $0$, a space-like singularity arises at $r\to 0$.}
\label{Bhxiao}
\end{figure}

Once we choose the $m=0$, the solution depicts a black hole or a horizonless de Sitter core depending on $\alpha$. When $0<\alpha\leq\frac{8 p^2}{243}$, the initial regular solution corresponds to a black hole with at most two horizons. As $m$ decreases from $0$, a timelike singularity appears at $r\to0$. The two horizons move closer together and eventually merge into one, forming an extremal black hole. If $m$ is decreased further, a naked timelike singularity develops. Conversely, when $m$ increases from $0$, the singularity at $r\to0$ becomes spacelike. For a slight increase in $m$, the black hole may exhibit nearly three horizons. As $m$ continues to increase, the two inner horizons merge, and for large $m$, these inner horizons disappear, leaving only an event horizon. These processes are illustrated in the left plot of Fig.\,\ref{Bhxiao}. 

The initial regular solution depicts a horizonless de Sitter core if $\alpha>\frac{8p^2}{243}$. When $m$ decreases from $0$, a naked timelike singularity always appears at $r\to0$. In other words, there are no black hole horizons when $m<0$ in this case. However, a spacelike singularity emerges at $r\to0$ if $m>0$. At least one horizon encloses the singularity. The metric function is not monotonically increasing outside the single horizon. As $m$ increases further, an extremal outer horizon forms. Continuing to increase $m$, a black hole with three horizons appears. The inner two horizons eventually merge as $m$ continues to grow. For large $m$, only a single horizon remains. Unlike the single-horizon black hole at small $m$, the large $m$ case resembles a Schwarzschild black hole, with a metric function that increases monotonically outside the event horizon. These processes are illustrated in the right plot of Fig.\ref{Bhxiao}.

\subsection{The singular counterparts satisfy the black hole first law}

The thermodynamic quantities of black holes, such as temperature and entropy, can be read as 
\be
T=\ft{f'(r_h)}{4\pi},\qquad S= \pi r_h^2,
\ee
where $r_h$ denotes the event (outer) horizon.

As a well-known result, the Bardeen black hole does not satisfy the first law of black hole thermodynamics. The reason is again that these regular black holes are just the special cases of full solutions, where their mass and charge are both fixed. 

In contrast, the two integration constants $(m,p)$ of the full solutions \eqref{fullsol} are free. The behavior of $r\to\infty$ of \eqref{fullsol} is shown in \ref{Bhxiao}.
It shows that the magnetic charge $p$ in the mass expression, namely the black hole mass now is 
\be
M=\ft{m}{2}+ \ft{2^{\fft{1}{4}}p^{\fft{3}{2}}}{3  \alpha^{\fft{1}{4}}}.
\ee
The first law of the black hole thermodynamics is 
\be \label{firstlaw}
d M=T dS+\phi_m dp,
\ee
where the magnetic potential $\phi_m=\ft{\sqrt{p}}{2^{\fft{3}{4}}\alpha^{\fft{1}{4}}}$ can get from \eqref{firstlaw}
\be
\phi_m=-\int^\infty_0 \ft{p}{r^2}\ft{\partial {\cal F}}{\partial F^2}dr,
\ee
and the temperature is 
\be
T=\ft{1}{4\pi r_h}(1-\ft{2^{\fft{11}{4}}p^{\fft{5}{2}}r_h^2\alpha^{\fft{1}{4}}}{(r_h^2+2\sqrt{2\alpha}p)^\fft{5}{2}}).
\ee

\subsection{Discontinuity of black hole temperature and entropy}

There are two different behaviours of the thermodynamic quantities, depending on the initial regular solution are regular black holes or horozionless de Sitter cores.

\subsubsection*{\centering A. Deviation from the black hole states}

\begin{figure}[h]
\centering
\includegraphics[width=0.45\textwidth]{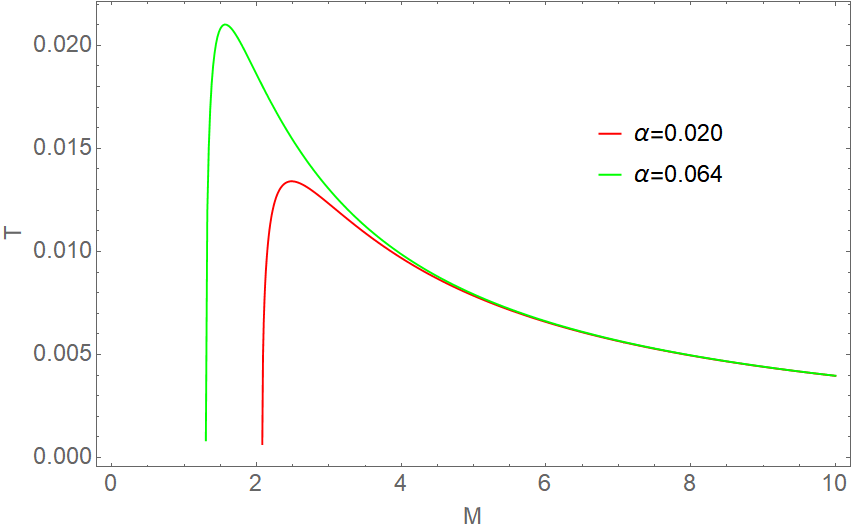} \quad
\includegraphics[width=0.45\textwidth]{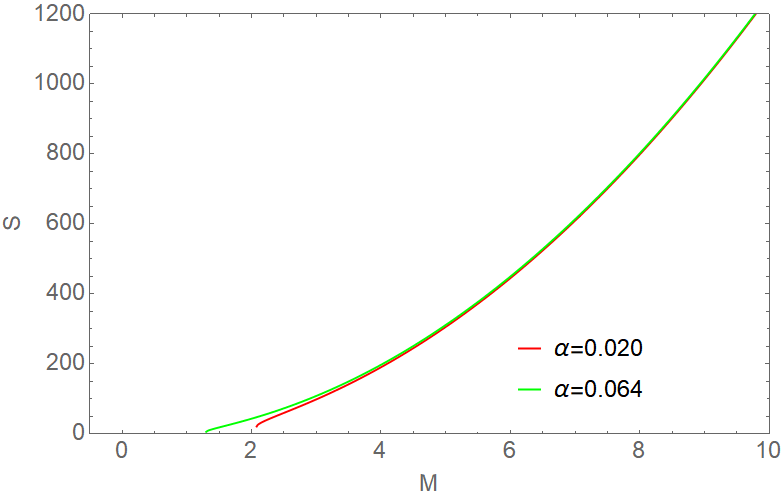} 
\caption{ \it  The thermodynamic temperature (left) and entropy (right) with varying black hole mass $M$ of the Bardeen black holes. Here we set $p=\ft{1}{2\sqrt{2\alpha}}$. For $\alpha=0.020$ and $\alpha=0.064$, the initial state is a two-horizons regular black hole and an extremal regular black hole, respectively. }
\label{bar}
\end{figure}

For initial states that are regular black holes, whether they are extremal or not, they share similar properties in thermodynamics. In small mass $M$, the temperature increases as $M$ increases. When the initial state is an extremal black hole, increasing $M$ leads it to be a non-extremal black hole. It should be noted that the temperature is undefined for the horizonless de Sitter core. On the other hand, when the initial state is not an extremal black hole, reducing $M$ will lead it to be extremal and hence has a zero temperature. Both cases have a maximum of temperatures $T=T_{max}$. Then in large $M$, the temperature decreases as $M$ increases. We present an example in the left plot of Fig.\ref{bar}. 

The entropy (area) of the black holes increases monotonically as $M$ increases. We present an example in the right plot of Fig.\ref{bar}.

\subsubsection*{\centering B. Deviation from the horizonless de Sitter core state}

There is no black hole thermodynamics if the initial state is a horizonless de Sitter core. However, with increasing $M$, the de Sitter core becomes a black hole. Then we can investigate the black hole thermodynamics.

For small $m>0$\footnote{It means $M>\ft{2^{\fft{1}{4}}p^{\fft{3}{2}}}{3  \alpha^{\fft{1}{4}}}$.}, an event horizon arises and it closes to the space-like singularity at $r\to 0$. Furthermore, the black hole radius grows as $M$ increases and hence the entropy also increases. The black hole temperature $T$ decreases as $M$ increases.

However, extra horizons will suddenly appear due to the fact that the metric function $f(r)$  is no longer monotonically increasing once $M>M_{cr}$, where $M_{cr}$ is the value that makes the outer horizon an extremal horizon. This is like a transition that a small black hole becomes a big black hole. Even though black hole entropy $S$ as a function of $M$ is still monotonically increasing, it has a discontinuity at $M=M_{cr}$.  The black hole temperature $T$ also has a discontinuity as a function of $M$. Since the outer horizon becomes extremal at $M=M_{cr}$, the temperature goes to zero. After that, with the increase of $M>M_{cr}$, $T$ increases from $0$ to a certain value then decreases.

\begin{figure}[h]
\centering
\includegraphics[width=0.45\textwidth]{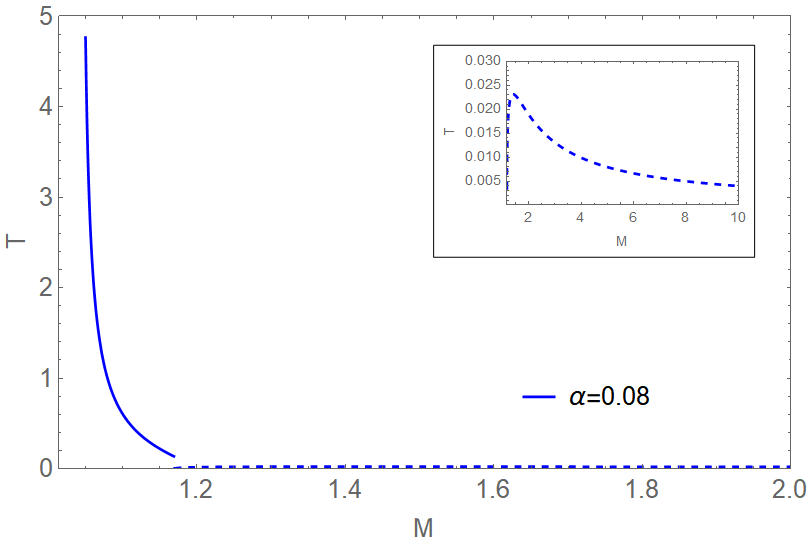}
\includegraphics[width=0.45\textwidth]{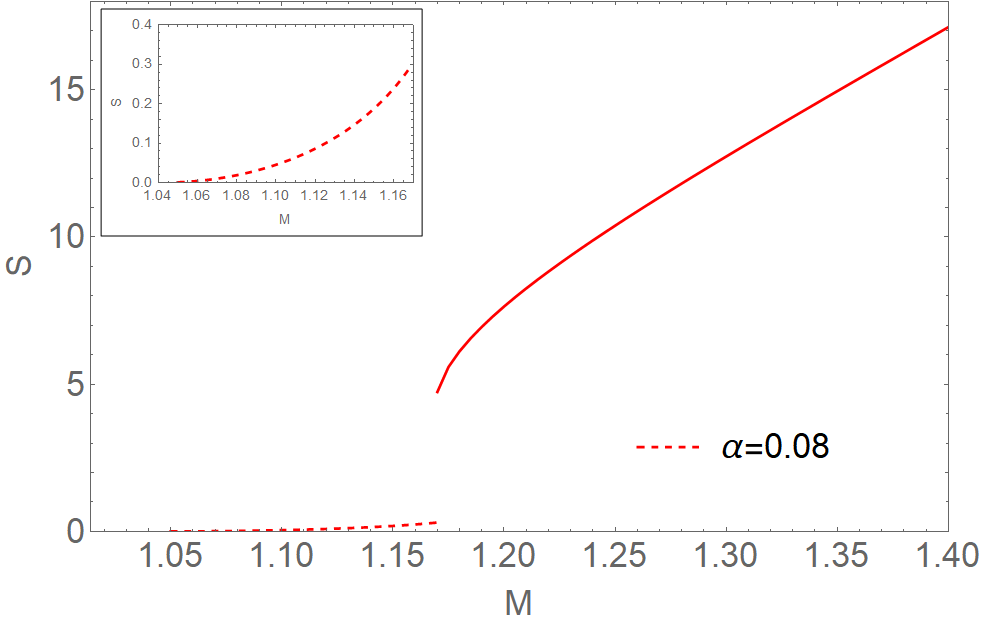} 
\caption{ \it The thermodynamic temperature (left) and entropy (right) with varying black hole mass $M$ of the Bardeen black holes. Here we set $p=\ft{1}{2\sqrt{2\alpha}}$. For $\alpha=0.08$, the original state is a de Sitter core. One can see there is a discontinuity at $M_{cr}=1.17
$, due to an outer horizon suddenly forms.
}
\label{barluo}
\end{figure}

These cases are illustrated in Fig.\ref{barluo}.

\section{Conclusion and Discussion}\label{5}

In this work, we investigated how regular black holes and singular black holes are related in General Relativity. We found that regular black holes are actually special cases, whereas more general solutions typically contain singularities.

In addition, we presented an example in NLED theory with four theoretical parameters. This theory can reduce to both the Bardeen class and the Hayward class. We obtained a new black hole solution in which regular black holes arise at a specific mass value.

We also examined differences in properties between singular and regular solutions, using the Bardeen black hole as an example. We found that while the regular solution does not contain any free parameters, the general solution includes a free parameter $m$ which relates to mass. By varying $m$, the black hole can have up to three horizons. Furthermore, we discovered that the metric function is not necessarily monotonically increasing outside the event horizon when the black hole has only one horizon — a behaviour contrasting with related regular black holes. This property suggests these black holes might produce gravitational wave echoes during the ringdown phase, which will be explored in future work.

When a small-mass black hole with a single horizon transitions into a multi-horizon black hole, we found that this process may lead to abrupt changes in the black hole’s entropy (i.e., its area) and temperature, causing them to no longer remain continuous functions of the mass. Consequently, such discontinuities in entropy and temperature with respect to mass might offer new insights for black hole chemistry \cite{Mann:2024sru,Kumar:2020cve,Hale:2025veb} and shadow images \cite{Gralla:2019xty,Huang:2024bbs}.

Furthermore, while standard regular black hole solutions do not strictly satisfy the first law of black hole thermodynamics, the inclusion of the full solution space restores thermodynamic consistency. These results highlight the fine-tuned nature of regular black hole models and suggest that singularity-free configurations are not generic in Einstein gravity.

It should be noted that our work is just a first glimpse into this topic, considering only spherically symmetric spacetime. Generalizations to broader cases remain subjects for future studies.

\section*{Acknowledgement}

We thank Hong L\"u for offering the original idea and his kind guidance. This work was supported by the National Natural Science Foundation of China (NSFC) Grant No. 12205123 and Jiangxi Provincial Natural Science Foundation with Grant No. 20232BAB211029.

\end{document}